\documentclass[trackchanges,resetfootnote]{aastex701}
\usepackage{mathrsfs}
\usepackage{graphicx}
\usepackage{comment}
\usepackage{amsmath}
\usepackage{psfrag}
\usepackage{multirow}%
\usepackage{amssymb,amsfonts}%
\usepackage{amsthm}%
\usepackage[title]{appendix}%
\usepackage{xcolor}%

\begin{document}

\title[Correlations of Simulated Black-Hole Movies]{
Extreme-Lensing Signatures Revealed by Correlations of Simulated Black-Hole Movies}

\author[orcid=0000-0002-8599-4483]{Barbora Bezd\v{e}kov\'{a}}
\affiliation{Department of Physics, Faculty of Natural Sciences, University of Haifa, Haifa 3498838, Israel}
\affiliation{Haifa Research Center for Theoretical Physics and Astrophysics, University of Haifa, Haifa 3498838, Israel}
\email[show]{bbezdeko@campus.haifa.ac.il}

\author[orcid=0000-0002-6960-0704]{Shahar Hadar}
\affiliation{Department of Physics, Faculty of Natural Sciences, University of Haifa, Haifa 3498838, Israel}
\affiliation{Haifa Research Center for Theoretical Physics and Astrophysics, University of Haifa, Haifa 3498838, Israel}
\affiliation{Department of Mathematics and Physics, University of
Haifa at Oranim, Kiryat Tivon 3600600, Israel}
\email{shaharhadar@sci.haifa.ac.il}

\author[orcid=0000-0001-6952-2147]{George Wong}
\affiliation{School of Natural Sciences, Institute for Advanced Study, Princeton, NJ 08540, USA}
\affiliation{Princeton Gravity Initiative, Princeton University, Princeton, NJ 08544, USA}
\email{gnwong@ias.edu}

\author[orcid=0000-0002-8635-4242]{Maciek Wielgus}
\affiliation{Instituto de Astrofísica de Andalucía, Gta. de la Astronomía, 18008 Granada, Spain}
\email{maciek.wielgus@gmail.com}

\begin{abstract}
A black hole's gravitational pull can deflect light rays to an arbitrary degree. As a result, any source fluctuation near the black hole creates multiple lagged images on an observer's screen. For optically thin stochastic emission, these light echoes give rise to correlations of brightness fluctuations across time-dependent images (movies). The correlation pattern disentangles source-specific characteristics from universal features dictated by general relativity. This picture has motivated a proposal to use the two-point image correlation function as a probe of extreme gravitational lensing in upcoming black-hole imaging campaigns. In this work, we test the feasibility of this method by computing the two-point correlation function of brightness fluctuations in a black-hole movie of state-of-the-art realism. The movie is generated by ray tracing a general relativistic magnetohydrodynamic simulation, which can then be blurred to any angular resolution. At an effective resolution expected to be achieved by next-generation terrestrial very-long-baseline interferometric arrays, these lensing signatures appear in neither time-averaged images nor light-curve autocorrelations. However, we demonstrate that they are clearly visible in the more fine-grained two-point image correlation function. Our positive findings motivate a more comprehensive investigation into the instrument specifications and inference techniques needed to resolve extreme lensing effects through correlations.
\end{abstract}


\section{Introduction}

By capturing the first resolved black hole (BH) images, the Event Horizon Telescope (EHT) has transformed our ability to investigate strong gravitational fields and the accretion flows they shape \citep{EHT19I,EHT19IV,EHT19V,EHT19VI,EHT22I,EHT22III,EHT22IV,EHT22V}.
Although the effect of gravity on light has been studied for more than a century and served as a main test of general relativity shortly after its publication \citep{dyson20},
imaging at horizon-scale resolution provides a new independent way to measure the warped spacetime geometry of a BH \citep[cf.][]{EHT22VI}.

A BH image is governed by two ingredients: the accretion flow into the hole, which emits the light; and the curved spacetime geometry, which lenses it.
The plasma forming the accretion flow typically exhibits complex behavior and is challenging to model; the current state of the art is set by general relativistic magnetohydrodynamics (GRMHD) numerical simulations. On the other hand, light propagation around a rotating (Kerr) BH is an analytically tractable, better understood problem \citep{Carter1968,bardeen73,gralla20}.
Thus, a key outstanding question emerges: how to most efficiently disentangle the complex astrophysics from the simple general relativistic effects in BH images?

Recent work has revealed that a universal, largely astrophysics-insensitive part of the image is the photon ring: a quasi-circular brightness enhancement in optically thin BH images, that results from extreme gravitational lensing. Note that we use the term ``extreme lensing'' to describe deflection of light by large angles, of the order of a radian or more; ``strong lensing'' often refers to situations with multi-path propagation, but small deflection angles. The photon ring arises from indirect light rays, that execute a number $n>0$ of half-orbits around the BH before traveling to the telescope, and appear exponentially close to a closed \textit{critical curve} on the observer's screen \citep{bardeen73};
this curve is the locus of directions whose light rays, traced backward, asymptotically approach a bound photon orbit. Equivalently, it is the image-plane separatrix between null geodesics that reach infinity and those that do not.
Thus, the photon ring is a self-similar stack of increasingly demagnified images of the emission.
Since the critical curve and demagnification pattern are determined by BH mass, distance, spin, and inclination angle relative to the line of sight,
the photon ring is governed primarily by the spacetime geometry, with only weak dependence on astrophysical details
\citep[e.g.,][]{Gralla2019,Johnson2020,Wielgus2021,  palumbo22,tiede22,Vincent2022,jia24,Urso2025}.
However, a difficulty arises; the photon ring is too thin to be resolved in time-averaged images with terrestrial very-long-baseline interferometry (VLBI) instruments, such as the EHT \citep{Lockhart2022,paugnat22,tiede22}.
This has recently motivated the planning of space VLBI missions to spatially resolve the photon ring, such as the Terahertz Exploration and Zooming-in for Astrophysics \citep[THEZA;][]{Gurvits2022,Hudson2023} and the Black Hole Explorer  \citep[BHEX;][]{BHEX2024}.

A desirable complementary approach would be to develop alternative techniques to probe extreme gravitational lensing without spatially resolving the photon ring.
Exploiting the time dependence of images potentially offers such an approach: the temporal lags between direct ($n=0$) and indirect photons with different $n>0$ induce correlations in time-dependent BH images
\citep[``movies'';][]{fukumura08,moriyama19,chesler21,hadar21,wong21,andrianov22,hadar23,Kocherlakota2024,wong24,harikesh25}.
In particular, \cite{hadar21} proposed the two-point correlation function of intensity fluctuations across different times and positions in a BH movie as an observable that could potentially reveal extreme lensing effects even when the photon ring is unresolved in time-averaged images.
Questions such as whether or not this method is realistically applicable, and which correlation function probes extreme lensing most effectively, were left open.

In this work, we take a step towards addressing them and analyze the two-point correlation function of intensity fluctuations in a BH movie simulated from a high-resolution, high-fidelity GRMHD simulation imaged using general relativistic ray tracing.
Our results indicate that
the most fine-grained image correlation function (without any integration over the two-point configuration space) reveals
the extreme-lensing signatures, even with realistic resolution of (next-generation) terrestrial-VLBI missions such as the next-generation Event Horizon Telescope \citep[ngEHT;][]{johnson23,Doeleman2023,Ayzenberg2025}.
Although there are still several open questions to be further examined, our analysis provides compelling evidence that extreme lensing effects may be detectable via correlation measurements, not only \emph{in principle}, but also \emph{in practice}.

\section{Methods}
\subsection{Simulated Black Hole Movie}
We analyze BH movies generated by ray tracing a 3D GRMHD simulation of a magnetically arrested disk (MAD) evolved with {\tt{iharm3d}} at a resolution of $384\times192\times192$ initialized with the standard MAD setup described in \cite{wong21}.
The GRMHD fluid snapshots were saved at a cadence of $0.1\, GM/c^3$, where $G$ is Newton's constant and $c$ is the speed of light.
Radiative transfer is performed with {\tt{ipole}} \citep{moscibrodzka18}, assuming an ideal fluid and a thermal electron distribution with a prescribed ion-to-electron temperature ratio parameterized by $R_{\rm low} = 1$ and $R_{\rm high} = 40$ \citep{Moscibrodzka2016}.
Images are generated at an observing frequency of 230 GHz for a Kerr BH with mass $M=6.5\times10^9 M_\odot$, dimensionless spin $a_*=0.9375$, and inclination angle between the spin axis and line of sight $\theta_o=163^\circ$, at a distance of D = 16.9 Mpc (having $GM/c^2 = 3.8 \,\mu$as and mimicking the parameters of the M87* BH)
with a total duration of $4000 \, GM/c^3$ and with images produced every $0.5 \, GM/c^3$. These parameters are consistent with the EHT modeling results for the M87* source \citep{EHT19V,EHT22V}.
The GRMHD evolution assumes ideal (non-radiative) GRMHD. The radiative transfer includes only thermal synchrotron emission and absorption with a prescribed ion-to-electron temperature ratio. Non-thermal particle populations or heating, Compton scattering, and radiative cooling/feedback are not included. Since the correlation effect we search for arises from geometric multi-path propagation and associated light-travel-time delays, additional plasma microphysics is expected primarily to modify the emissivity distribution and thus the amplitude of the signal, rather than qualitatively remove the lensing-induced correlation feature in the optically thin regime considered here.

We used three distinct ray-tracing prescriptions to generate the movies. The main results of the paper are derived from the realistic \textit{slow-light} prescription, in which the state of the fluid is allowed to evolve as light propagates throughout the simulation domain.
In addition to this realistic ray-tracing calculation, we also used two auxiliary prescriptions to help interpret the correlations in the main movie.
In the first prescription, light is only emitted along the segments labeled $n=0$ following the convention defined below, so that the images do not contain signatures of photons that have been lensed around the far side of the BH. The literature adopts multiple conventions for labeling subring/segment order $n$ (e.g., time-of-flight ordering versus geometric segmentation; see \cite{wong24,zhou25} for discussion and comparison), and consequently subring-separated images depend on the convention adopted.
In this work, we define segment boundaries at turning points of $dz/ds$, that is, when the geodesic reverses its vertical direction of propagation and begins heading back toward the $z=0$ midplane.
By construction, this method includes most of the \textit{astrophysical correlations}, arising from local interaction in the source, and excludes \textit{(gravitational-)lensing correlations}, which arise from light multi-path propagation.
The second auxiliary method uses the common \textit{fast-light} prescription, which calculates the image for a given time using a single simulation snapshot, as if the light had infinite speed, suppressing the temporal correlation structure due to lensing effects. In particular, we note that the EHT libraries of images, used for comparisons between observations and theory, are constructed using the fast-light approximation \citep{EHT19V,EHT22V}.
For more details on slow- and fast-light simulations, see, for example, \cite{wong22,conroy23}.

An additional aspect of the movies we generated is the associated image resolution.
As an initial test of the applicability of our method to upcoming finite-resolution real data, which would be gathered by future instruments such as ngEHT or BHEX,
we blurred the images by different Gaussian kernels characterized by their full width at half maximum (FWHM).
To illustrate the effect of different blurring kernels, time-averaged images obtained from the blurred slow-light movie are shown in %Supplementary
Fig.~\ref{fig_ex_blurr}. The expected nominal (diffraction-limited) angular resolution of ngEHT is $\sim15 \, \mu$as \citep{johnson23}, while that of BHEX is estimated at $\sim6 \, \mu$as \citep{BHEX2024,sridharan24}. The maximum resolution enabled by the super-resolution image reconstruction algorithms is often estimated at $\sim50\%$ of that, hence $\sim3 \, \mu$as for BHEX.

\subsection{Two-Point Image Correlation Function} \label{sec:2PT image correlation}
The main object of study in this work is the (normalized) two-point correlation function
\begin{equation}\label{corr_fun_6D}
\mathcal{C}(T,x,y,x',y')=\frac{1}{\sigma\left[\Delta I(t,x,y)\right] \sigma\left[\Delta I(t,x',y')\right]}\langle \Delta I(t,x,y)\Delta  I(t-T,x',y')\rangle,
\end{equation}
of intensity fluctuations $\Delta I(t,x,y)=I(t,x,y)-\langle I(t,x,y) \rangle$ in a BH movie, where $(t,x,y)$ parameterize the observer time and Cartesian, Bardeen screen coordinates \citep{bardeen73}, respectively, $\sigma\left[\Delta I(t,x,y)\right]$ denotes the temporal standard deviation of the intensity at screen position $(x,y)$, and $\langle \rangle$ denotes time averaging.
Notice that if $I(x,y)=\mathrm{const.}$ throughout the simulation, $\sigma(x,y)=0$ and Eq.~\eqref{corr_fun_6D} is not defined. However, in the present simulation this issue does not occur.
The time lag is taken to be positive: $T\geq0$, that is, the point at $(x',y')$ is taken at an earlier time compared to the point $(x,y)$. Generally, $\mathcal{C}$ is not invariant under time reversal. Time-reversal symmetry is severely broken by the accretion inflow. The Kerr background also breaks this symmetry, but supplementing it with an axial reflection yields the discrete isometry $(t,\phi) \to  (-t,-\phi)$.
We used here a normalized version of $\mathcal{C}$, slightly modifying the original non-normalized definition in \cite{hadar21}. On the coincidence 2D surface $\{T=0,x=x',y=y'\}$, by definition, $\mathcal{C}(0,x,y,x,y)=1$.

As argued in \cite{hadar21}, in an optically thin setting, the function $\mathcal{C}(T,x,y,x',y')$ provides potential means to disentangle non-universal (astrophysical) and universal (gravitational lensing) information about the source, as follows.
Gravitational-lensing correlations reflect the relationship between photons that are produced at the same spacetime event in the source, but undergo a different number of half orbits $n$. Indeed,
such photons generally travel along distinct paths and reach the observer at different screen positions and times. The $n \geq 1$ images converge rapidly---exponentially in $n$---to the critical curve, appearing at different angles around it. Regarding the time lag, for (even moderately) small inclination such as in the case considered here, the typical time lag between images of consecutive $n$ is $\approx 15\,GM/c^3$ \citep[e.g.,][]{gralla20lensing, Kocherlakota2024, wong24}, and this value depends only weakly on BH spin.
As two relevant examples, we note that for M87*, $15\,G M/c^3 \approx 5$ days, while for Sgr~A*, $15\,G M/c^3 \approx 5$ minutes.
Naturally, the most significant contribution to the image (in terms of the observed flux density) comes from the direct ($n=0$) image and the first extremely lensed image ($n=1$), and therefore the most significant contribution to lensing correlations originates from the correlation of the $n=0,1$ contributions.
If source correlations are local, astrophysical correlations are expected to decay monotonously away from their maximal value, attained at the coincidence surface $\{T=0, x=x', y=y'\}$. Therefore, if they exist, \textit{isolated} local maxima of correlation found away from the coincidence surface are a smoking-gun signature of gravitational-lensing correlations.

Since $\mathcal{C}(T,x,y,x',y')$ lives in a 5D configuration space, visualizing its detailed structure is challenging. At least two possible strategies may be adopted to this end: coarse graining by integrating over some directions of the configuration space, or investigating it on a slice-by-slice basis, for instance by fixing some of the coordinates and studying correlation maps on the resulting subspace. A couple of particular proposals for coarse graining via integration were discussed in \cite{hadar21}; these seem to be inapplicable in the present case, as they are too coarse; the lensing correlation enhancements become obscured by astrophysical correlations. Nevertheless, coarse-grained observables can be useful for overcoming observational limitations \citep[e.g.,][]{hadar23}. On the other hand, if the observing coverage is sufficient to allow for a robust reconstruction of instantaneous snapshot images, as seems to be the case for upcoming extensions of EHT, the fine-grained correlation function $\mathcal{C}(T,x,y,x',y')$ is always superior to coarse-grained alternatives in terms of the information it conveys.
For example, the most coarse-grained observable that is still sensitive to time lags is the (normalized) two-point correlation function of light curve fluctuations, given by
\begin{equation}\label{corr_fun_1D}
\mathcal{C}_\mathrm{LC}(T)=\frac{1}{\sigma^2\left[\Delta F(t)\right]}\langle \Delta F(t)\Delta  F(t+T)\rangle,
\end{equation}
where $F(t)=\int\int I(t,x,y)dx\,dy$ defines the light curve, or total flux density in the image.
This observable is used broadly in many different types of astronomical observations.
\cite{wielgus22} and \cite{cardenas24} argued in detail why BH extreme lensing signatures cannot be found in $\mathcal{C}_\mathrm{LC}$.
The essential reason, as mentioned above, is that projecting the function via integration onto a significantly lower-dimensional configuration space mixes different correlation contributions, and the astrophysical correlations dominate the gravitational-lensing contributions.

\begin{figure*}[ht]
\centering
\includegraphics[width=.65\textwidth]{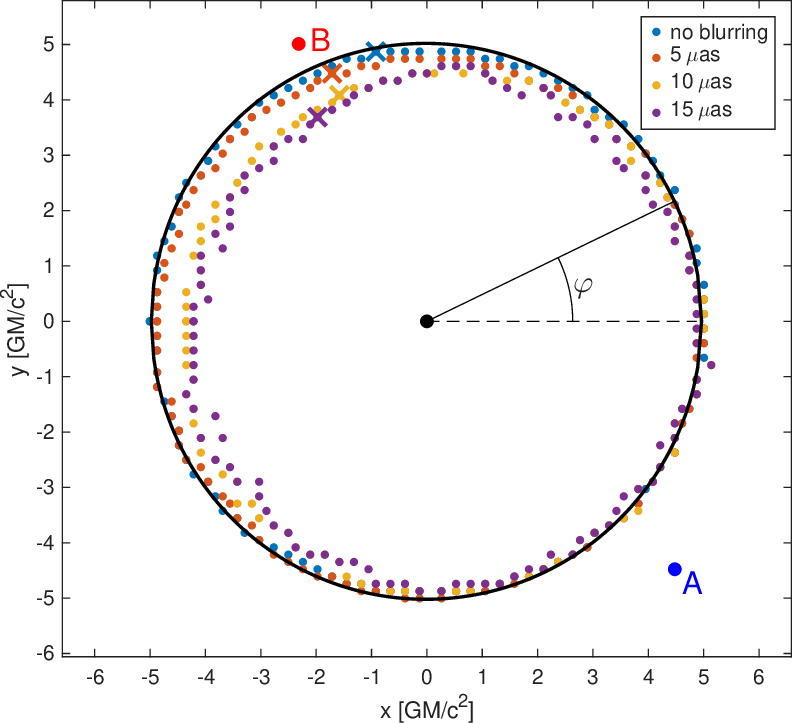}
\caption{Empirical and theoretical critical curves along with the positions of our fixed point and its secondary image. The empirical critical curves were derived for movies of different blurrings (color-coded dots) and the (exact) theoretical critical curve is the solid black. The blue point, A, marks the position of our fixed point, set at $(4.346,-4.610)~GM/c^2$, and the red point, B, shows the theoretically predicted position of its secondary image (assuming an equatorial source). The color crosses show points of highest correlation located on the given empirical critical curve. The black point is our chosen pole, set at $(0,0)~GM/c^2$.}\label{points_cc} %$(25,33)~\mu \mbox{arcsec}$
\end{figure*}

In this paper, we take the second approach and visualize our results for $\mathcal{C}$ by restricting to particular subspaces of the 5D configuration space.
To begin with, we fix the screen coordinates of the earlier point to $(x_0',y_0')=(4.346,-4.610)~GM/c^2$, the blue point A in Fig.~\ref{points_cc}.
It is important to emphasize that this choice is rather arbitrary and was based on esthetical reasons for presentation purposes. It is made to reduce the dimensionality of the configuration space (to 3 dimensions) and simplify the analysis. In doing so, we use only a small portion of the information contained in \eqref{corr_fun_6D}. We remark that a similar analysis with the same qualitative results was performed also for different choices of $(x_0',y_0')$, see Sec.~\ref{sec:results}.

In addition, we reduce by one more dimension in two different ways:
\begin{itemize}
\item We find the closed curve of highest intensity in its time-averaged image, which we term the ``empirical critical curve'' $\rho_c(\varphi)$, where $(\rho,\varphi)$ are polar coordinates on the screen, and the pole is chosen to be the centroid of $\rho_c(\varphi)$ of the unblurred case, at $(x,y)_\mathrm{pole} \approx (0,0)~GM/c^2$. In practice, we found the position of a brightest pixel along a ray originating at $(x,y)_\mathrm{pole}$ with angular increments of $\pi/55$ rad between rays.
We have verified that shifting the pole position by $\sim \mathrm{few} \, \mu$as (a substantial fraction of $GM/c^2$) changes the correlation maps negligibly, so our results are rather robust to the choice of pole.
Fig.~\ref{points_cc} shows the empirical critical curves we determine for movies of different resolutions. This reduction to 2D yields a correlation map $\mathcal{C}(T,\rho_c(\varphi),\varphi,x_0',y_0')$, visualizing the correlation structure in the time/azimuth plane (Fig.~\ref{fig_corr_maps}).

\item We present the correlation on fixed $T=T_0$ surfaces, that is, $\mathcal{C}(T_0,x,y,x_0',y_0')$ as a function of the screen coordinates of the late point $(x,y)$; Fig.~\ref{fig_corr_T}. Our results for the correlation maps are discussed in the next section.

\item Additionally, we present the correlation structure off the critical curve in the time/azimuth plane, displaying $\mathcal{C}(T,\rho_c(\varphi)+\Delta \rho,\varphi,x_0',y_0')$, where $\Delta \rho$ is the deviation off the curve (Fig.~\ref{fig_corr_maps_rs}).
\end{itemize}

\section{Results} \label{sec:results}

\begin{figure*}[ht!]
\centering
\includegraphics[width=1\textwidth]{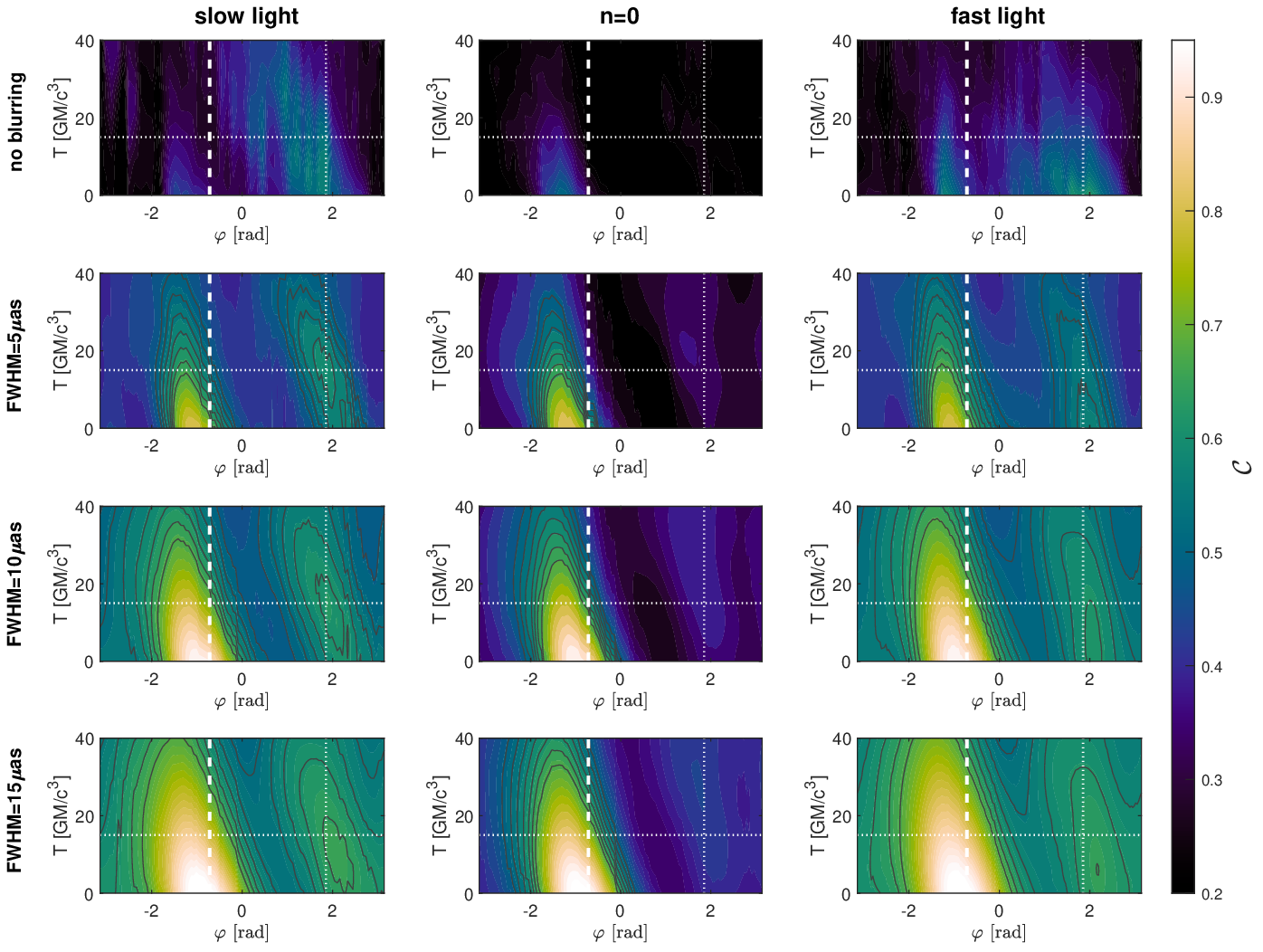}
\caption{Two-point correlation function of intensity fluctuations
as a function of the angular position $\varphi$ and the time lag $T$. The angle $\varphi$ is measured along the empirical critical curve.
The correlation function $\mathcal{C}(T,\rho_c(\varphi),\varphi,x_0',y_0')$ (see Eq.~\ref{corr_fun_6D}) was obtained with different blurring kernels (rows) and ray-tracing prescriptions (columns). The dashed line shows the angular position $\varphi_0'=-\pi/4 \, \mathrm{rad}$ of the point A, that we fix in this analysis, located at $(x_0',y_0')=(4.346,-4.610)~GM/c^2$, and the horizontal and vertical dotted lines mark the expected positions of the secondary peak in the temporal and angular domains, respectively. The contour levels plotted in gray correspond to values 0.5, 0.53, 0.56, 0.59, 0.62, 0.65, 0.68.}\label{fig_corr_maps}
\end{figure*}

The correlation maps $\mathcal{C}(T,\rho_c(\varphi),\varphi,x_0',y_0')$, for different values of blurring kernels (rows) and ray-tracing prescriptions (columns) are shown in Fig.~\ref{fig_corr_maps}.

In individual panels, two significant lobes of correlation enhancement are apparent. The first lobe peaks near the coincidence limit: at $T=0 \, GM/c^3$, and close to the angular position of the fixed point A, given by $\varphi_0' = -\frac{\pi}{4} \, \mathrm{rad}$ (dashed vertical line). This enhancement is strongly dominated by the astrophysical correlations;
indeed, for given blurring it is approximately similar in magnitude and shape in all ray-tracing prescriptions. However, the second lobe receives significant contributions from lensing correlations;
it occurs around $T=15\,GM/c^3$, the characteristic half-orbit time delay for (moderately) small inclinations, and its magnitude and shape are very sensitive to the ray-tracing scheme used. The dotted lines correspond to the approximate temporal lag and angular position ($T \approx 15 \, GM/c^3$ and $\varphi \approx 1.85 \,\mathrm{rad}$, respectively) expected for the lensing correlation peak. The angular position of the secondary image for point A was calculated by adapting the method proposed by \cite{zhou25}, assuming an equatorial source.

With slow-light ray tracing, the position of the peak of the lensing correlation lobe is in good agreement with the theoretical expectations for the $n=1$ image. In the fast-light case, the lensing lobe still occurs, but is significantly modified as its peak shifts towards $T=0 \, GM/c^3$, which is a natural consequence of the scheme's insensitivity to time delays. In the direct-only ($n=0$) case, the lensing lobe is diminished to varying degrees, depending on the width of the blurring kernel; it almost completely fades away with no blurring. The correlation at the position of the lensing lobe does not completely fade away at finite blurring since as the blurring kernel becomes wider, non-lensing (astrophysical) correlation from the $n=0$ image of the flow orbiting the BH appears close enough to the later point to affect it through blurring. In all cases, it is evident that the lensing peak is suppressed compared to the astrophysical one, but it is still possible to distinguish the two since the lensing lobe peaks at a local maximum away from the coincidence point.
While
the astrophysical correlation peak, at $\varphi \approx -\pi/4 \, \mathrm{rad}$, is very similar in both the slow-light and direct-only ray-tracing prescriptions (including its maximal value), the magnitude and shape of the lensing correlation peak, at $\varphi \approx 1.8 \, \mathrm{rad}$, differs significantly between ray-tracing prescriptions. In particular, in the slow-light case its maximal value is
significantly higher than in the direct-only case (their ratio varies significantly with blurring, a factor of $\sim2$ at $10 \, \mu$as).

As explained in Sec.~\ref{sec:2PT image correlation}, a key indication of lensing correlations is that they form isolated peaks in the configuration space. The peaks seen in $(\varphi,T)$-space in Fig.~\ref{fig_corr_maps} are encouraging, but do not conclusively imply that we have identified an isolated lensing lobe (a local maximum in the configuration space which is separated from the coincidence point, and is a smoking-gun signature of extreme lensing)
since they could in principle still be connected in the full 5D configuration space.
While we leave an exhaustive investigation of the isolation of lensing correlations in the 5D space to future work, here we present evidence of their isolation in the 3D configuration subspace defined by fixing the early point to a particular (arbitrary) position, $(x',y')=(x_0',y_0')$.
We only present results for the choice $(x_0',y_0')=(4.346,-4.610)~GM/c^2$ (point A in Fig.~\ref{points_cc}), but we have checked several other choices for $(x_0',y_0')$ that show similar behavior; their detailed analysis will be presented in our future work. Nevertheless, a few additional examples with different choices of $(x_0',y_0')$  are shown in
%Supplementary
Figs.~\ref{first supplementary map}-\ref{last supplementary map}.

\begin{figure*}[ht!]
\centering
\includegraphics[width=1\textwidth]{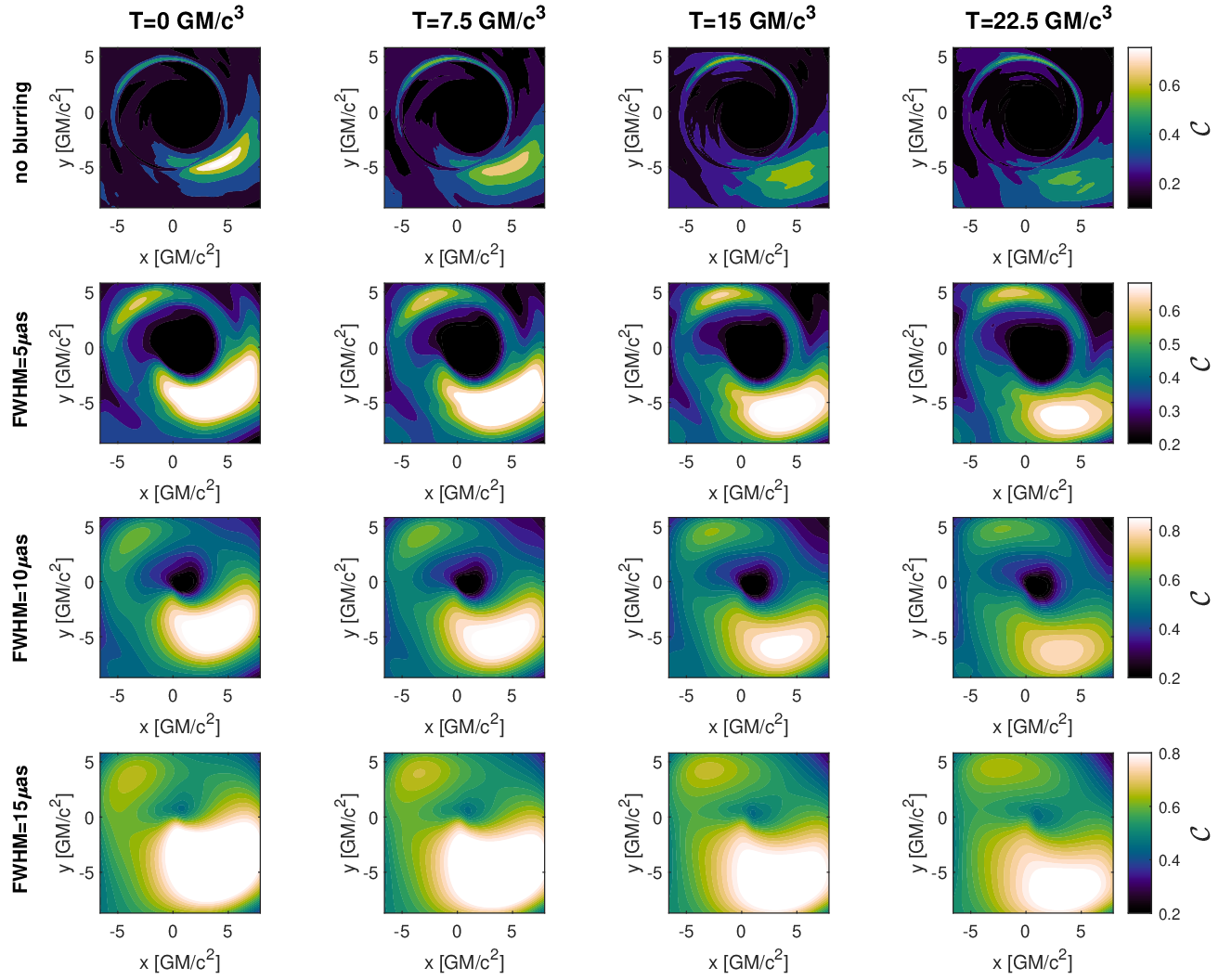}
\caption{Correlation maps $\mathcal{C}(T_0,x,y,x_0',y_0')$ for different choices of time lag $T_0$ and blurring kernel widths. Each map was generated from slow-light ray-traced movies. The astrophysical correlation lobe peaks near the bottom right corner of the image at $T=0 \, GM/c^3$, while the lensing correlation lobe peaks near the upper left corner of the image around $T=15 \, GM/c^3$. These correlation maps are consistent with the presence of a local maximum of $\mathcal{C}(T,x,y,x_0',y_0')$ in the 3-dimensional configuration space $\{T,x,y\}$, constituting a signature of extreme lensing.}\label{fig_corr_T}
\end{figure*}

To analyze the correlation structure in the 3D subspace,
we take two different approaches.
One is to fix surfaces of particular time lag, $T=T_0$, and plot $\mathcal{C}(T_0,x,y,x_0',y_0')$.
Such correlation maps,
derived from slow-light movies, are shown in Fig.~\ref{fig_corr_T} for different time lags and blurring kernels. For each blurring, the correlations peak around the coincidence limit $(x,y)=(x_0,y_0)$ at $T=0 \, GM/c^3$, and this peak gradually diminishes with larger $T$. However, another correlation enhancement appears in a limited range of radii around $(x,y)=(-2,4.5)~GM/c^2$ (roughly on the opposite side of the critical curve); it is clearly isolated from the primary peak at fixed $T$.
The radial range of this secondary lobe is naturally extended with larger blurring, but it
remains separated from the primary peak. Moreover,
it
peaks at $T>0 \, GM/c^3$, typically around $15 \, GM/c^3$; note, however, that
for blurring with FWHM$=15 \, \mu \mbox{as}$, the maximum occurs already at lower $T$ (but is still well-separated from $T=0 \, GM/c^3$). This seems to arise from the smearing of correlations by
finite-resolution effects.
In the second approach, we study the correlation structure for deviations off the empirical critical curve.

\begin{figure*}[ht!]
\centering
\includegraphics[width=1\textwidth]{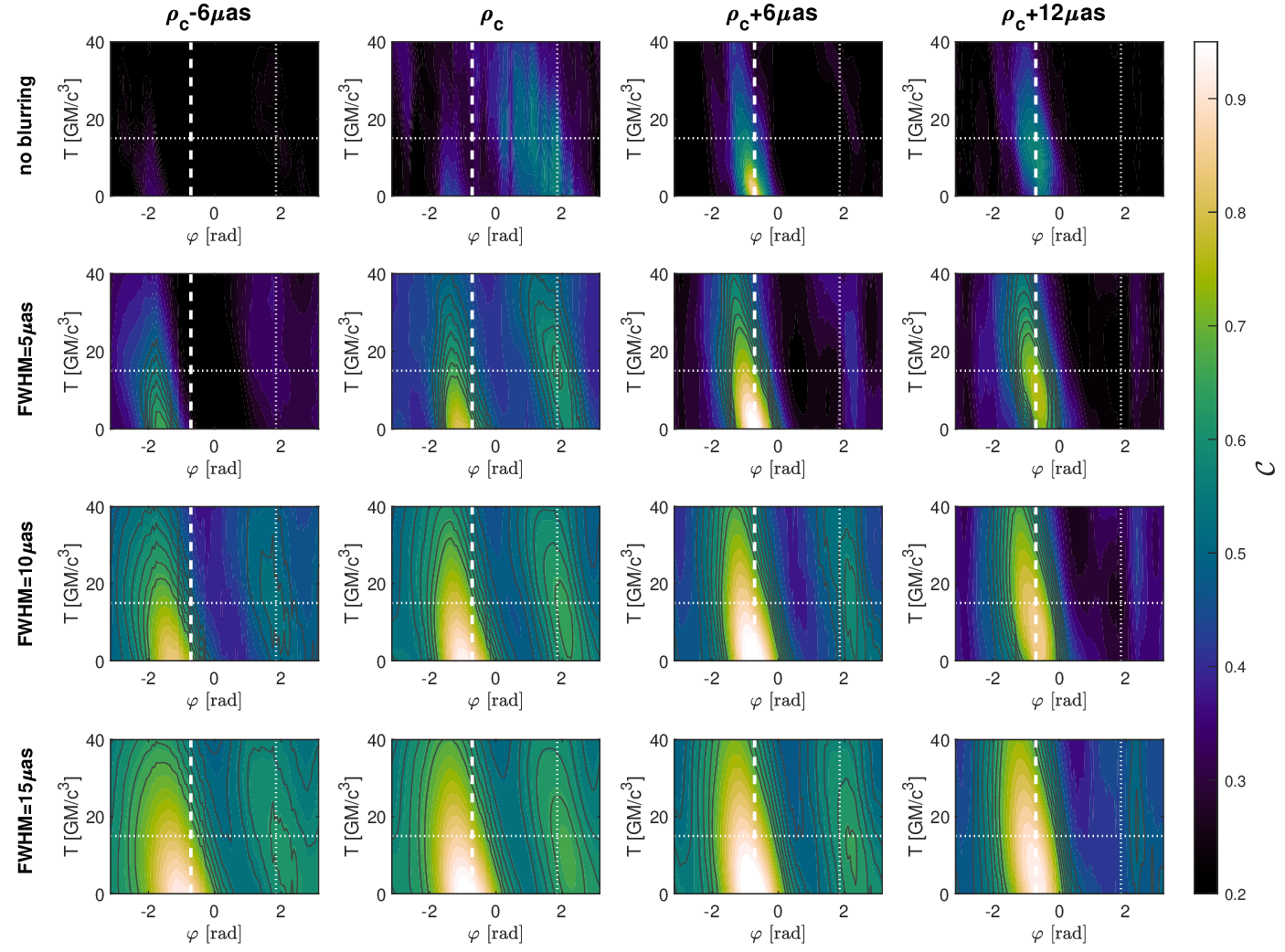}
\caption{Two-point correlation function of intensity fluctuations $\mathcal{C}(T,\rho_c(\varphi)+\Delta \rho,\varphi,x_0',y_0')$ as a function of the angular position $\varphi$ and the time lag $T$. The correlation function (see Eq.~\eqref{corr_fun_6D}) was derived for different values of $\Delta \rho=-6,0,6,12 \, \mu\mbox{as}$ (columns) and blurrings (rows).
The dashed line shows the angular position $\varphi_0'=-\pi/4 \, \mathrm{rad}$ of the point we fix in this analysis, located at point A (see Fig.~\ref{points_cc}), and the horizontal and vertical dotted lines mark the expected positions of the secondary peak in the temporal and angular domains, respectively.}\label{fig_corr_maps_rs}
\end{figure*}

The maps shown in Fig.~\ref{fig_corr_maps} were obtained by fixing the screen position of the early point at $(x_0',y_0')$ (reducing the full 5D configuration space to 3D), and constraining the late point $(x,y)$ to lie on the empirical critical curve $\rho=\rho_c(\varphi)$ (which reduces one additional dimension). Our results show two separated lobes of enhanced correlation, each of which has an isolated peak (local maximum). However, this fact does not necessarily imply that the correlation enhancements are separated in the 3D configuration space, where the late point is allowed to lie anywhere on the screen.
As a preliminary check that the lensing peak is indeed isolated, we visualized the reduced 3D configuration space (with the early point fixed) in two different ways. In the first one, shown in Fig.~\ref{fig_corr_T}, we display different 2D slices of fixed $T=T_0$. In the second, presented in Fig.~\ref{fig_corr_maps_rs}, we show different 2D slices of fixed $\rho=\rho_c(\varphi)+\Delta \rho$, with a deviation $\Delta \rho$ from the empirical critical curve.

Fig.~\ref{fig_corr_maps_rs} shows correlation maps in the $(\varphi,T)$-domain for four choices of $\Delta \rho=-6,0,6,12 \, \mu\mbox{as}$ (corresponding to $\approx-1.58, 0,    1.58,3.16 ~GM/c^2$)
(columns), generated from movies with four different blurring kernels with FWHM=$0,5,10,15 \, \mu \mbox{as}$ (rows).
The asymmetric choice of values for $\Delta \rho$ is natural since $\mathcal{C}$ falls off more rapidly toward the inside of the image than it does towards the outer side.

We examined such 2D correlation maps for many values of $\Delta \rho$. The correlation peaks fall off in all directions, including the radial one, and do so increasingly steeply with better resolution.
For all values of the blurring kernel FWHM, we found no evidence of a monotonically rising ``ridge''  connecting the lensing peak at $\Delta \rho = 0$ to the primary peak at $\Delta \rho = 0$. This provides further evidence that the lensing peak is isolated, i.e., that it is a local maximum of $\mathcal{C}$ in the full 3D configuration space.

\begin{figure*}[ht!]
\centering
\includegraphics[width=1\textwidth]{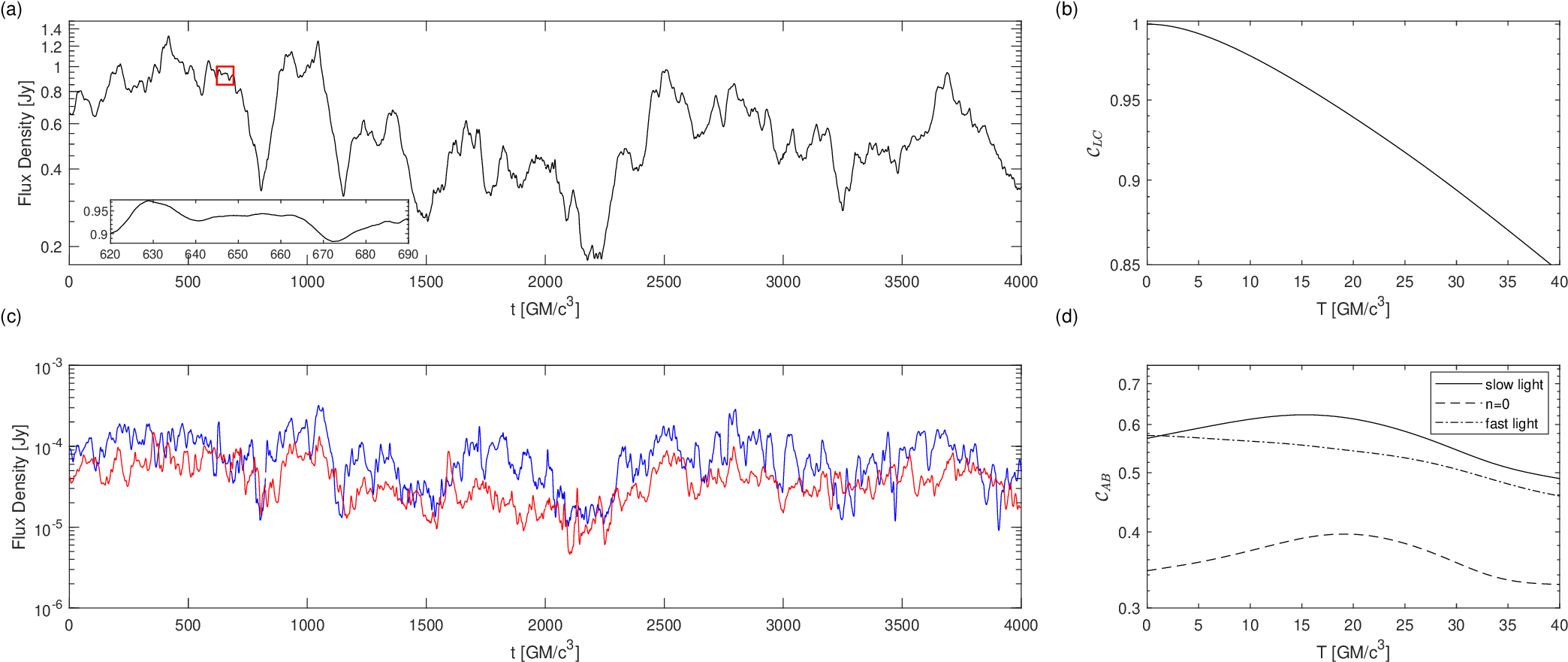}
\caption{Examples of the 1D observables derived from our data sets. (a) Light curve of the slow-light black-hole movie. The red rectangle shows the position of the zoomed-in light curve depicted in the bottom subpanel. (b) Autocorrelation function of the light curve shown in panel (a).  (c) Flux density profiles obtained from the slow-light movie, corresponding to pixels at points A (blue) and B (red); see Fig.~\ref{points_cc}. (d) Cross-correlation functions of the flux densities at points A, B derived from the slow-light (solid line), $n=0$ (dashed line), and fast-light (dashed-dotted line) movies.} \label{fig_corr_lc}
\end{figure*}

In order to connect and compare with recent work \citep{wielgus22,cardenas24} on light-curve autocorrelations, we extracted the light curves from the slow-light movies and computed their autocorrelation function $\mathcal{C}_\mathrm{LC}(T)$, see Eq.~\ref{corr_fun_1D}. The results are presented in the upper panels of
Fig.~\ref{fig_corr_lc}.
While the light curve fluctuates significantly,
even at the characteristic lensing timescale $\sim15 \,GM/c^3$
(see the subpanel in Fig.~\ref{fig_corr_lc}a), $\mathcal{C}_\mathrm{LC}(T)$ decreases monotonically with $T$ from $T=0 \, GM/c^3$ (where it is maximal, by definition) and does not show a peak around $T=15 \, GM/c^3$ (Fig.~\ref{fig_corr_lc}b).
In contrast, in Fig.~\ref{fig_corr_lc}c we plot the (slow-light, FWHM$=5 \, \mu \mbox{as}$) profiles of flux density corresponding to square pixels of size $0.5~\mu$as ($0.13~ GM/c^2$) at points A and B, the primary and secondary image locations for the same equatorial source, marked in Fig.~\ref{points_cc} with blue and red dots, respectively. In Fig.~\ref{fig_corr_lc}d we compute their cross-correlation function
\begin{equation}\label{eq:cross_corr_fun}
\mathcal{C}_\mathrm{AB}(T)=\frac{1}{\sigma\left[\Delta F_A(t)\right]\sigma\left[\Delta F_B(t)\right]}\langle \Delta F_A(t)\Delta  F_B(t+T)\rangle,
\end{equation}
and find a lagged correlation peak roughly at the lensing timescale $\sim15 \,GM/c^3$. We also plot similar cross-correlation functions extracted from the direct-only and fast-light movies. We find that the direct-only cross-correlation function is significantly suppressed in magnitude. In addition, we find a difference in shapes: for the direct-only one we identify a shifted peak at $\sim20 \,GM/c^3$, and for the fast-light one we see that the cross-correlation function monotonously decreases. We find these results to be consistent with the correlation maps in Figs.~\ref{fig_corr_maps},\ref{fig_corr_T}.

\section{Discussion and Conclusions}
We have analyzed the two-point correlation function of intensity fluctuations in a GRMHD-simulated BH movie,
defined in Eq.~\eqref{corr_fun_6D}. This function lives on a 5D configuration space, but by (arbitrarily) fixing the screen coordinates of one of the points,
we reduced the configuration space
to a 3D subspace, and this allowed us to efficiently visualize the correlation patterns
and analyze their dependence on lag and screen position of the second point.
Our results suggest that it may be possible to detect signatures of extreme lensing using correlations, even at the angular resolution of $\approx 15 \, \mu \mbox{as}$ expected in (next-generation) terrestrial VLBI missions---namely, with the ngEHT---for the mass-to-distance ratio $M/D$ of M87*. For Sgr~A* the $M/D$ is larger by about 25\%, reducing the resolution requirements, but on the other hand Sgr~A* suffers from the impact of the scattering screen, absent for M87*. In either case, the necessary angular resolution is achievable with ground VLBI arrays, possibly with a mild enhancement from algorithms enabling super-resolution. However, we stress that this statement should be taken as compelling evidence rather than a conclusive assertion, since we have only checked the effect of finite resolution on one simulation; to be more decisive, further realism must be introduced to this study, as we detail below.

A key challenge in detecting extreme-lensing effects through correlations
is to reliably show that the lensing-correlation enhancements are distinct from the astrophysical correlation peak.
Our results demonstrate that non-detection of the lensing peak in the (coarse-grained) light curve autocorrelation function $\mathcal{C}_\mathrm{LC}$, Eq.~\eqref{corr_fun_1D}, does not mean it cannot be detected in the (fine-grained) image correlation function $\mathcal{C}$, Eq.~\eqref{corr_fun_6D}. In other words, while previous studies reduced the dimensionality of configuration space by integrating over spatial directions, our results indicate that this is not the most efficient strategy, at least when the observing coverage is sufficient.
They clearly demonstrate the utility of fine-grained observables for revealing extreme-lensing correlations in the image domain, and show that coarse graining can lose essential information that obstructs the desired signatures.
See \cite{wong24} for a related demonstration of the advantage of fine-grained observables in the visibility domain.
A related disadvantage of coarse graining is that the resulting observables might be efficient only in certain regimes; for example, the radially-integrated (over both points' radii) correlation function is useful predominantly for (mildly) small inclinations. Let us emphasize that fine/coarse graining here refers to the correlation function, in particular to the dimensionality of the configuration space over which it is defined. A sufficiently high dimension (fine graining) of the configuration space appears to be crucial for detecting extreme lensing signatures.

In order to obtain a clearer picture of the instrumental specifications that could allow a measurement of lensing correlations,
we will need to improve the realism of the synthetic data we generate and test its detectability
in follow-up investigations. Some specific questions are:
a) The analysis in this study was performed in the image domain, while in reality VLBI images are reconstructed from incomplete Fourier domain sampling.
What is the realistic number of baselines required for our approach to be feasible?
b) EHT data might exhibit irregular temporal sampling due to calibration requirements. When does this fact become an impediment for correlation measurements?
c) At what level do instrumental, atmospheric, or other sources of noise obstruct correlation measurements?
d) What is the required observation time? Here we used a movie duration of 4000~$GM/c^3$, corresponding to $\approx22$ hours for Sgr~A* and $\approx3.7$ years for M87*; it seems entirely possible, however, that a significantly shorter time could be sufficient.
e) What is a good quantitative measure for detection? In other words, how can we estimate the prominence (or conversely the probability of false detection) of an extreme-lensing peak?
f) How does the correlation map change for fixed points in a different simulated movie?
We hope to address these aspects in a follow-up work, but we briefly comment on them below.

The ultimate detection criterion seems to be to show the separation/isolation of the lensing lobe in the 5D configuration space. Equivalently, one may work in the reduced 3D space with the early point fixed and show isolation for many different fixed positions of the early point. Since such an analysis is outside the scope of the present paper and as it is sometimes useful to have a simpler metric for the prominence of the lensing peak, we propose two criteria which (taken together) seem to be a good proxy of a prominent lensing peak. The first is that the value of (dimensionless) correlation at the peak be larger than some threshold value. A value of $0.5$ seems to be a good proxy in the present simulation. The second is the value of the Hessian determinant of the correlation function at the maximum, taken as a function over the 3D reduced configuration space (with the early point fixed). This quantity provides a measure of the second derivatives/curvature at the peak. We find that this quantity is roughly an order of magnitude larger for ``true'' lensing peaks appearing in the slow-light simulations than for the ``false'' lensing peaks appearing in the direct-only ($n=0$) simulations. We however stress that this proposal is preliminary, and will be further tested and elaborated upon in upcoming work.
From our ongoing analysis it is already clear that the total duration of the movie used in the current work (4000~$GM/c^3$) may be reduced and still allows a faithful measurement of correlations. Our findings suggest that a duration of $\sim$2000~$GM/c^3$ could be enough to detect extreme-lensing signatures for at least a decent subset of fixed points. We also note that the Hessian determinant stabilizes roughly for the same total duration as the correlation maps do. Furthermore, neither reduced temporal cadence nor temporal data segmentation (in a manner similar to that expected in future measurements such as ngEHT) seem to be obstacles for our technique.

Additionally, to make this technique a truly powerful tool,
a systematic approach for parameter inference should be developed.
How can the correlations be exploited to measure BH mass, spin, and inclination? How precise could such measurements be?
We will study these questions in detail in our future work. But already at this point, we have presented encouraging evidence in support of the conjecture that signatures of extreme lensing could be revealed with image correlations, even at a terrestrial-VLBI resolution. \\

%\bmhead{Data availability}
%Due to large amount of data used in this study, G.N.W. (gnwong@ias.edu) will provide the movies used in this study upon reasonable request, i.e., by an active researcher who is able to supply a means to transfer the data (e.g., a globus endpoint or ssh login credentials).

%\bmhead{Code availability}
%B.B. will provide the codes used to derive the results in this work upon request.

\begin{acknowledgments}
We are grateful to Michael Johnson for helpful comments. This work was supported in part by the Israel Science Foundation (grant No. 2047/23).
B.B. acknowledges support from the Simons Foundation (MP-SCMPS-00001470).
G.N.W. was supported by the Taplin Fellowship and the Princeton Gravity Initiative.
M.W. is supported by a Ramón y Cajal grant RYC2023-042988-I from the Spanish Ministry of Science and Innovation and acknowledges financial support from the Severo Ochoa grant CEX2021-001131-S funded by MCIN/AEI/ 10.13039/501100011033.
The computations which led to the results presented in this work were partially performed on the Hive computer cluster at the University of Haifa, which is partly funded by ISF grant 2155/15.
\end{acknowledgments}

\begin{contribution}
B.B. led the project. S.H. developed the original idea of detecting photon ring via correlations. B.B. and S.H. elaborated the technique how to handle the correlation function in a full generality (five-dimensional space). B.B. performed the analysis. G.N.W. generated data sets used in this work. M.W. delivered the data and provided comments to the analysis. B.B. and S.H. wrote the paper, G.N.W. and M.W. performed revisions of the final draft.
\end{contribution}

%\bmhead{Competing Interests}
%The authors declare no competing interests.

\appendix \label{sec:app supplementary figures}

%\subsection*{
\section{Supplementary material}

\subsection{Time-averaged images of data sets with different Gaussian blurring kernels}
\begin{figure*}[ht!]
\centering
\includegraphics[width=1\textwidth]{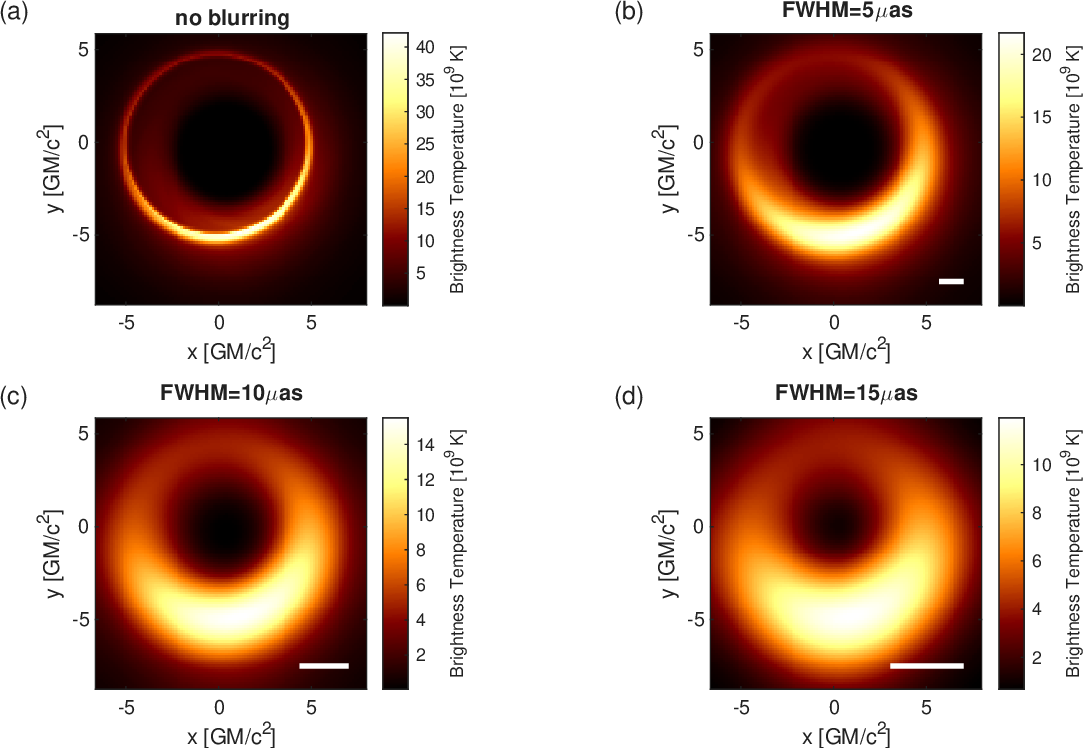}
\caption{Time-averaged images of intensity as a function of screen position, $\langle I(t,x,y) \rangle$, produced from simulated slow-light ray-traced movies. Panel (a) shows the unblurred image; other panels show time-averages of movies blurred by different Gaussian kernels, characterized by their FWHM, (b) $=5 \, \mu \mbox{as}$, (c) $=10 \, \mu \mbox{as}$, (d) $=15 \, \mu \mbox{as}$, to simulate the effect of limited instrumental resolution. Hereafter, for the assumed black hole parameters, mimicking those of M87*, $GM/c^2$ corresponds to $3.8\, \mu$as. The white horizontal lines in the bottom right corner of the blurred images show the FWHM of the applied kernels.}\label{fig_ex_blurr}
\end{figure*}

\newpage
\subsection{Examples of correlation maps obtained for additional fixed points $(x_0',y_0')$}

\begin{figure*}[ht!]
\centering
\includegraphics[width=1\textwidth]{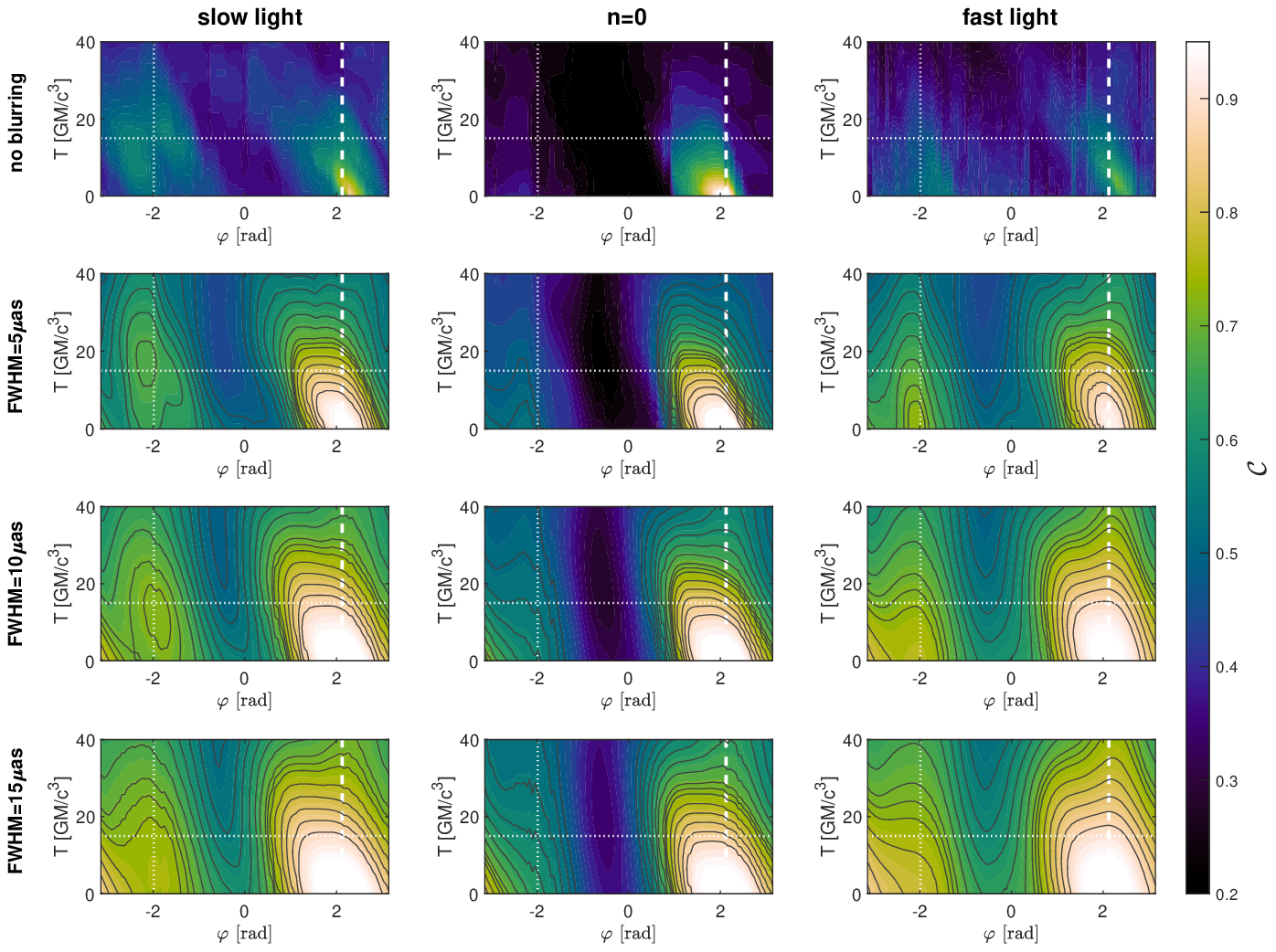} %corr_map_34_92
\caption{Two-point correlation function of intensity fluctuations
as a function of the angular position $\varphi$ and the time lag $T$.
The correlation function $\mathcal{C}(T,\rho_c(\varphi),\varphi,x_0',y_0')$ (see Eq.~1 in the Methods Section) was obtained with different blurring kernels (rows) and ray-tracing prescriptions (columns). The dashed line shows the angular position $\varphi_0'\approx 2\pi/3 \, \mathrm{rad}$ of the point which we fixed, located at $(x_0',y_0')=(-2.239, 3.293)~GM/c^2$, and the horizontal and vertical dotted lines mark the expected positions of the secondary peak in the temporal and angular domains, respectively. The contour levels plotted in gray correspond to values 0.5, 0.53, 0.56, 0.59, 0.62, 0.65, 0.68.}\label{first supplementary map}
\end{figure*}

\begin{figure*}[ht!]
\centering
\includegraphics[width=1\textwidth]{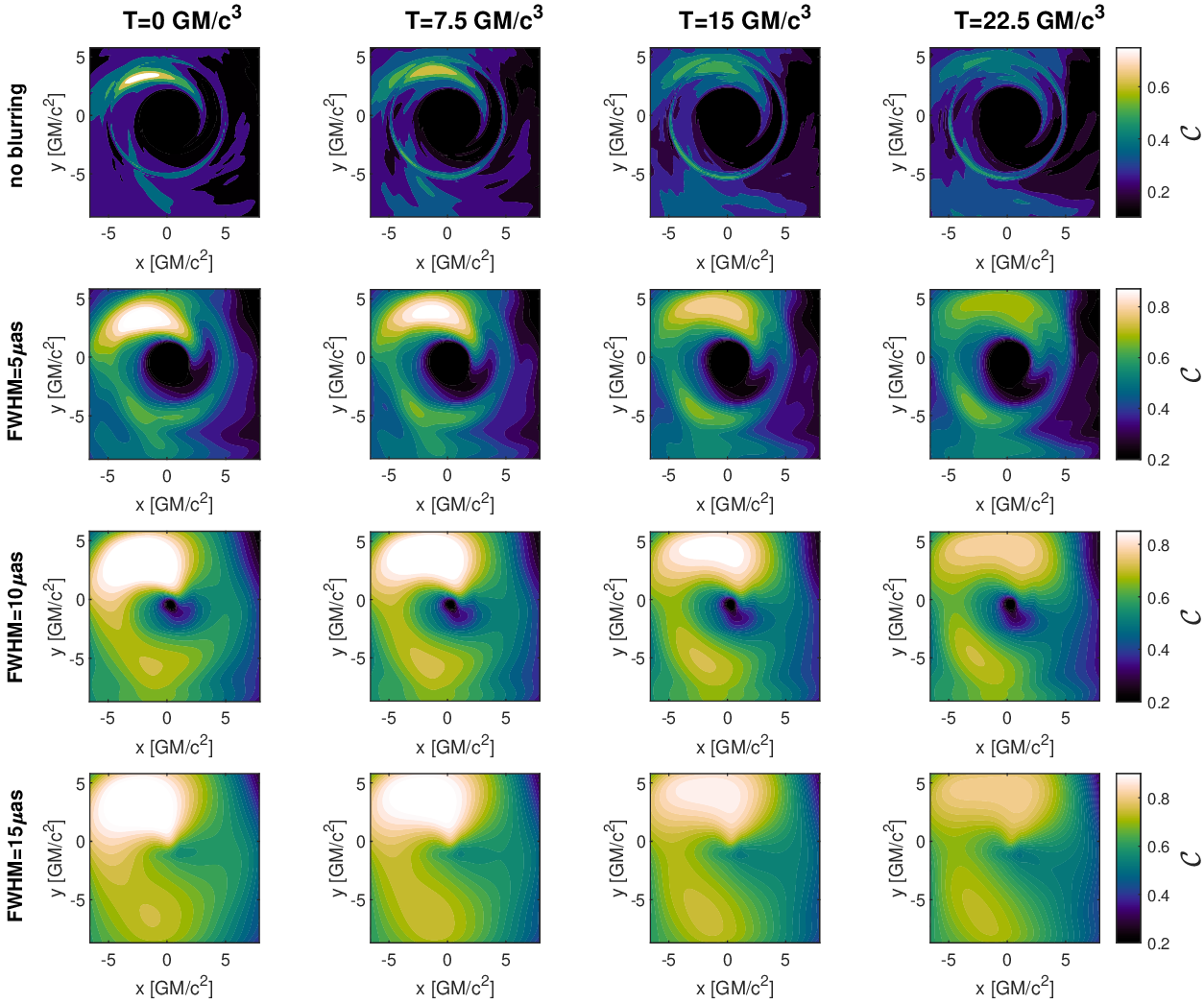} %corr_T_34_92
\caption{Correlation maps $\mathcal{C}(T_0,x,y,x_0',y_0')$ for different choices of time lag $T_0$ and blurring kernel widths for a fixed point at $(x_0',y_0')=(-2.239, 3.293)~GM/c^2$. Each map was generated from slow-light ray-traced movies. The astrophysical correlation lobe peaks near the upper left corner of the image at $T=0 \, GM/c^3$, while the lensing correlation lobe peaks near the bottom left corner of the image around $T=15 \, GM/c^3$.} % These correlation maps are consistent with the presence of a local maximum of $\mathcal{C}(T,x,y,x_0',y_0')$ in the 3-dimensional configuration space $\{T,x,y\}$, constituting a signature of extreme lensing.
\end{figure*}

\begin{figure*}[ht!]
\centering
\includegraphics[width=1\textwidth]{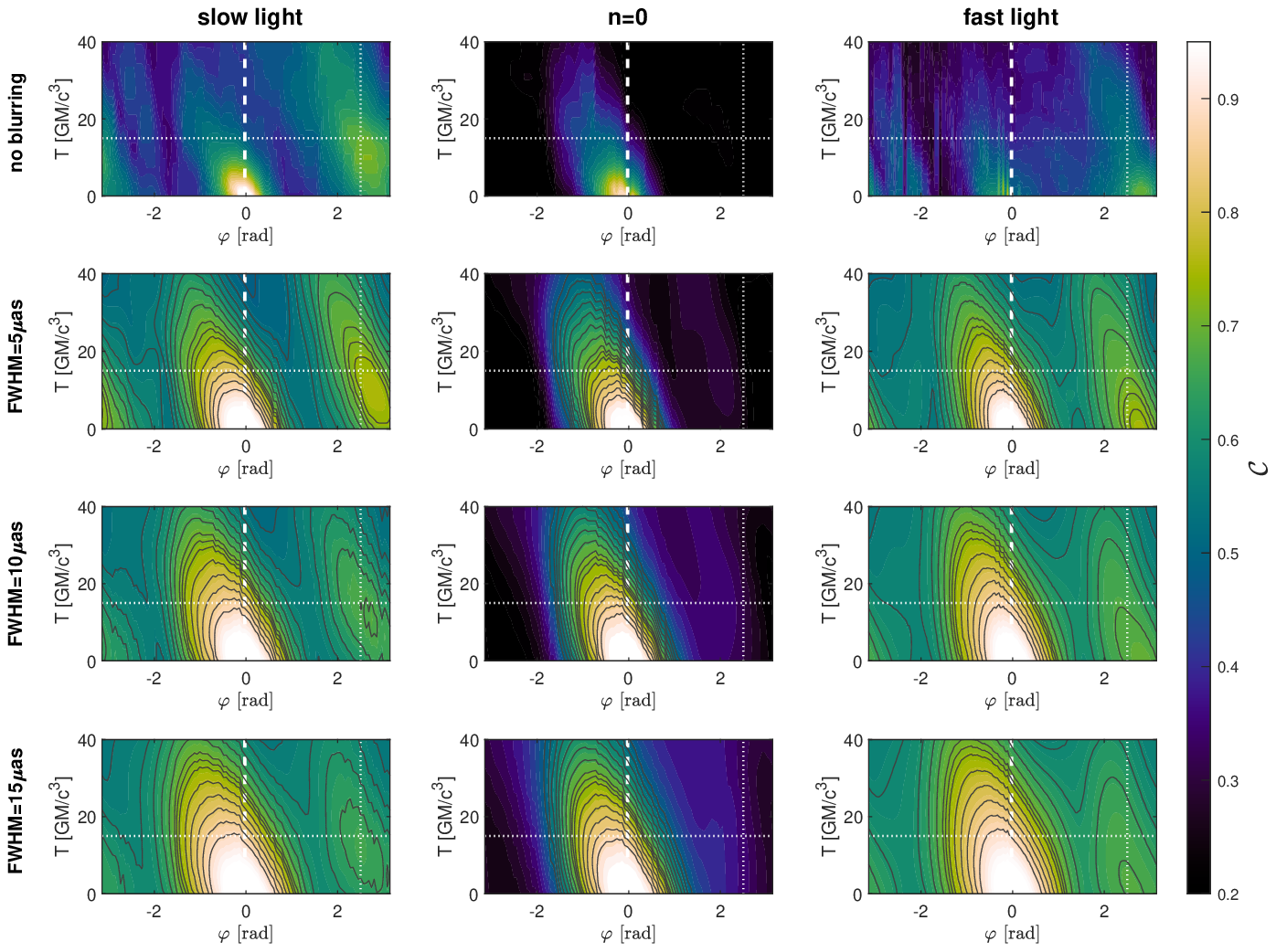} %corr_map_90_65
\caption{Two-point correlation function of intensity fluctuations
as a function of the angular position $\varphi$ and the time lag $T$.
The correlation function $\mathcal{C}(T,\rho_c(\varphi),\varphi,x_0',y_0')$ (see Eq.~1 in the Methods Section) was obtained with different blurring kernels (rows) and ray-tracing prescriptions (columns).
The dashed line shows the angular position $\varphi_0'=-0.025 \, \mathrm{rad}$ of the point which we fixed, located at $(x_0',y_0')=(5.136, -0.263)~GM/c^2$, and the horizontal and vertical dotted lines mark the expected positions of the secondary peak in the temporal and angular domains, respectively. The contour levels plotted in gray correspond to values 0.5, 0.53, 0.56, 0.59, 0.62, 0.65, 0.68.}
\end{figure*}

\begin{figure*}[ht!]
\centering
\includegraphics[width=1\textwidth]{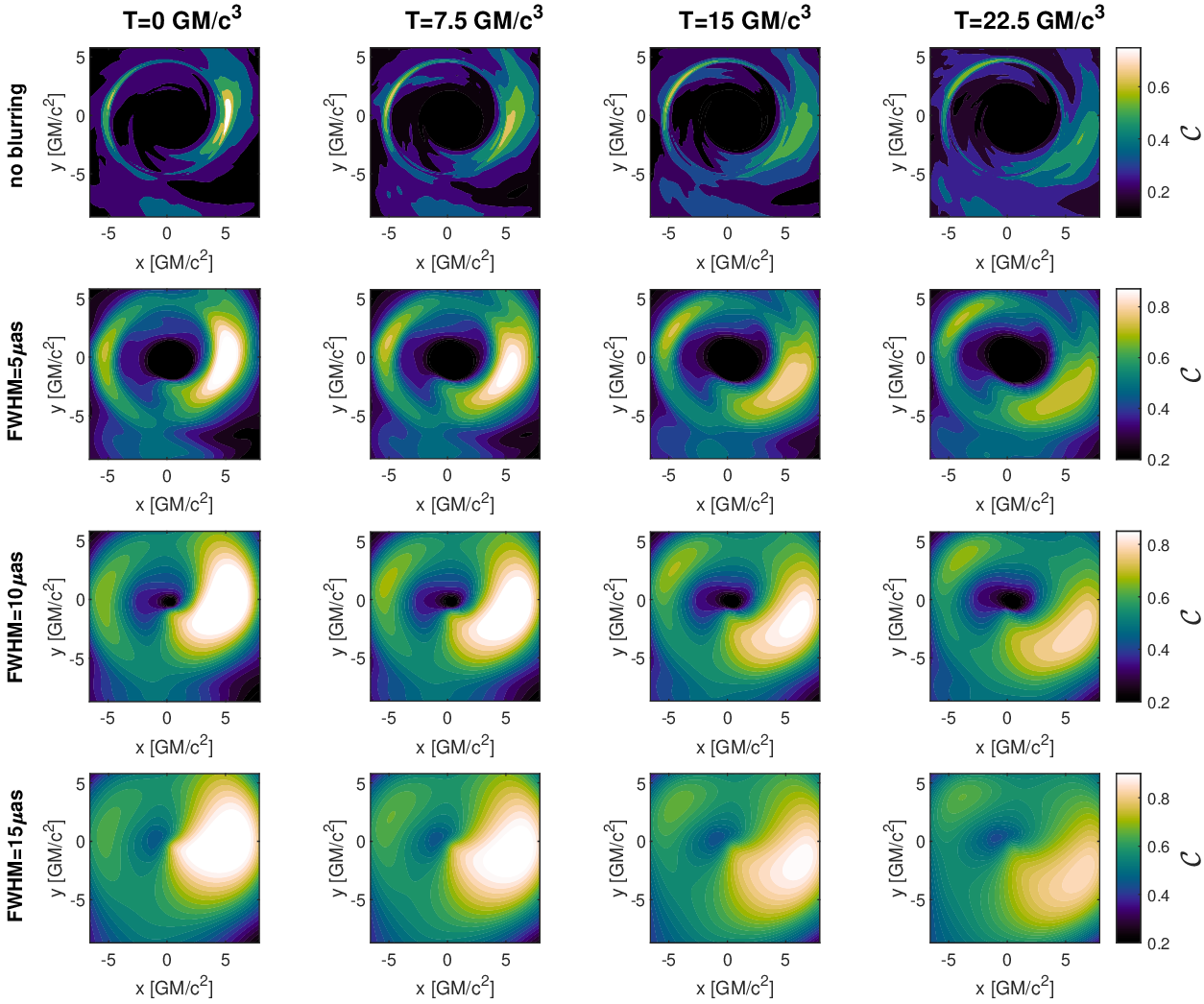} %corr_T_90_65
\caption{Correlation maps $\mathcal{C}(T_0,x,y,x_0',y_0')$ for different choices of time lag $T_0$ and blurring kernel widths for a fixed point at $(x_0',y_0')=(5.136, -0.263)~GM/c^2$. Each map was generated from slow-light ray-traced movies. The astrophysical correlation lobe peaks near the critical curve close to $\varphi=0 \, \mathrm{rad}$ of the image at $T=0 \, GM/c^3$, while the lensing correlation lobe peaks on the opposite side of the critical curve in the image around $T=15 \, GM/c^3$.} %These correlation maps are consistent with the presence of a local maximum of $\mathcal{C}(T,x,y,x_0',y_0')$ in the 3-dimensional configuration space $\{T,x,y\}$, constituting a signature of extreme lensing.
\end{figure*}

\begin{figure*}[ht!]
\centering
\includegraphics[width=1\textwidth]{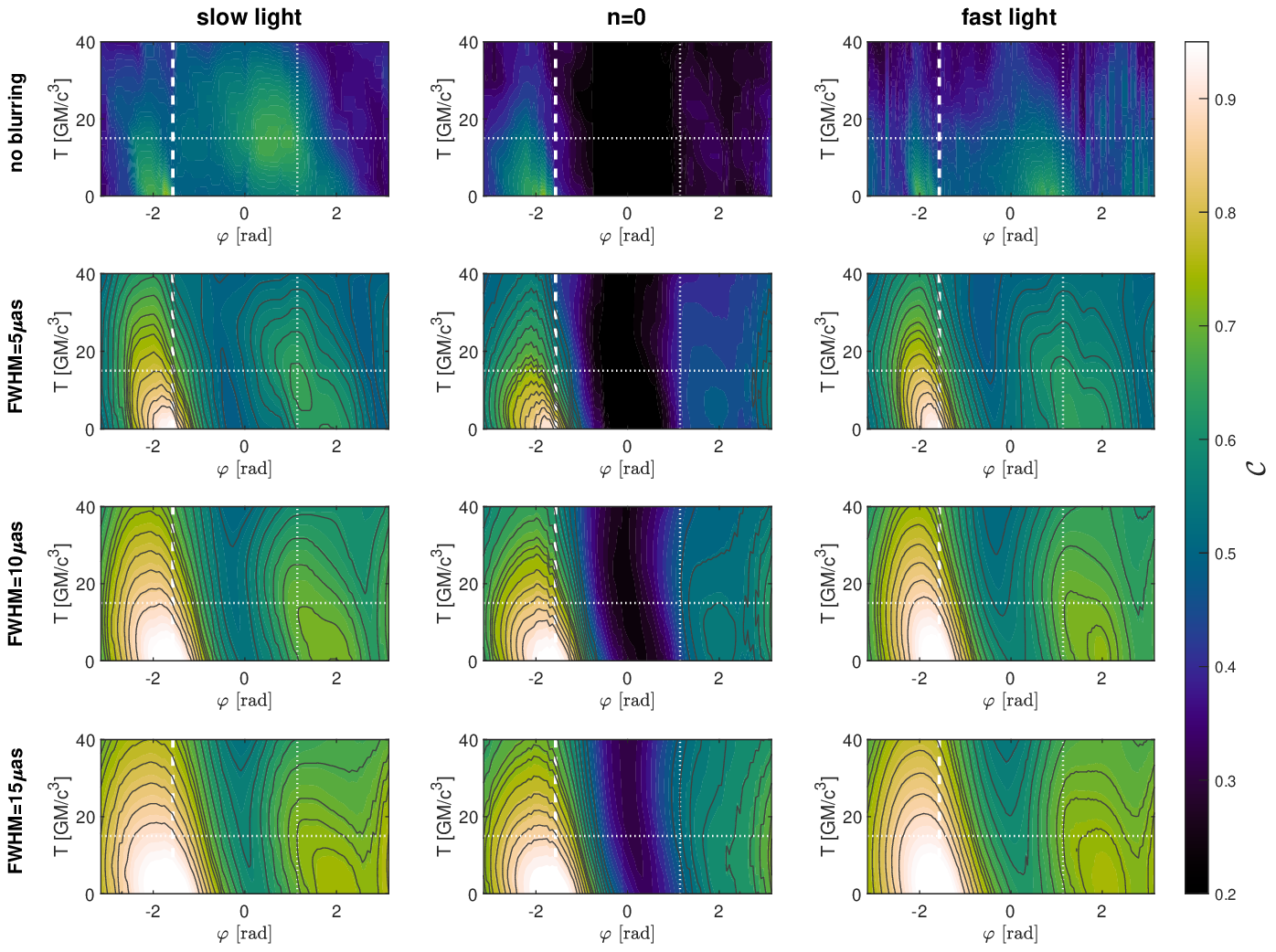} %corr_map_50_20
\caption{Two-point correlation function of intensity fluctuations
as a function of the angular position $\varphi$ and the time lag $T$.
The correlation function $\mathcal{C}(T,\rho_c(\varphi),\varphi,x_0',y_0')$ (see Eq.~1 in the Methods Section) was obtained with different blurring kernels (rows) and ray-tracing prescriptions (columns). The dashed line shows the angular position $\varphi_0'=-\pi/2 \, \mathrm{rad}$ of the point which we fixed, located at $(x_0',y_0')=(-0.132, -6.190)~GM/c^2$, and the horizontal and vertical dotted lines mark the expected positions of the secondary peak in the temporal and angular domains, respectively. The contour levels plotted in gray correspond to values 0.5, 0.53, 0.56, 0.59, 0.62, 0.65, 0.68.}
\end{figure*}

\begin{figure*}[ht!]
\centering
\includegraphics[width=1\textwidth]{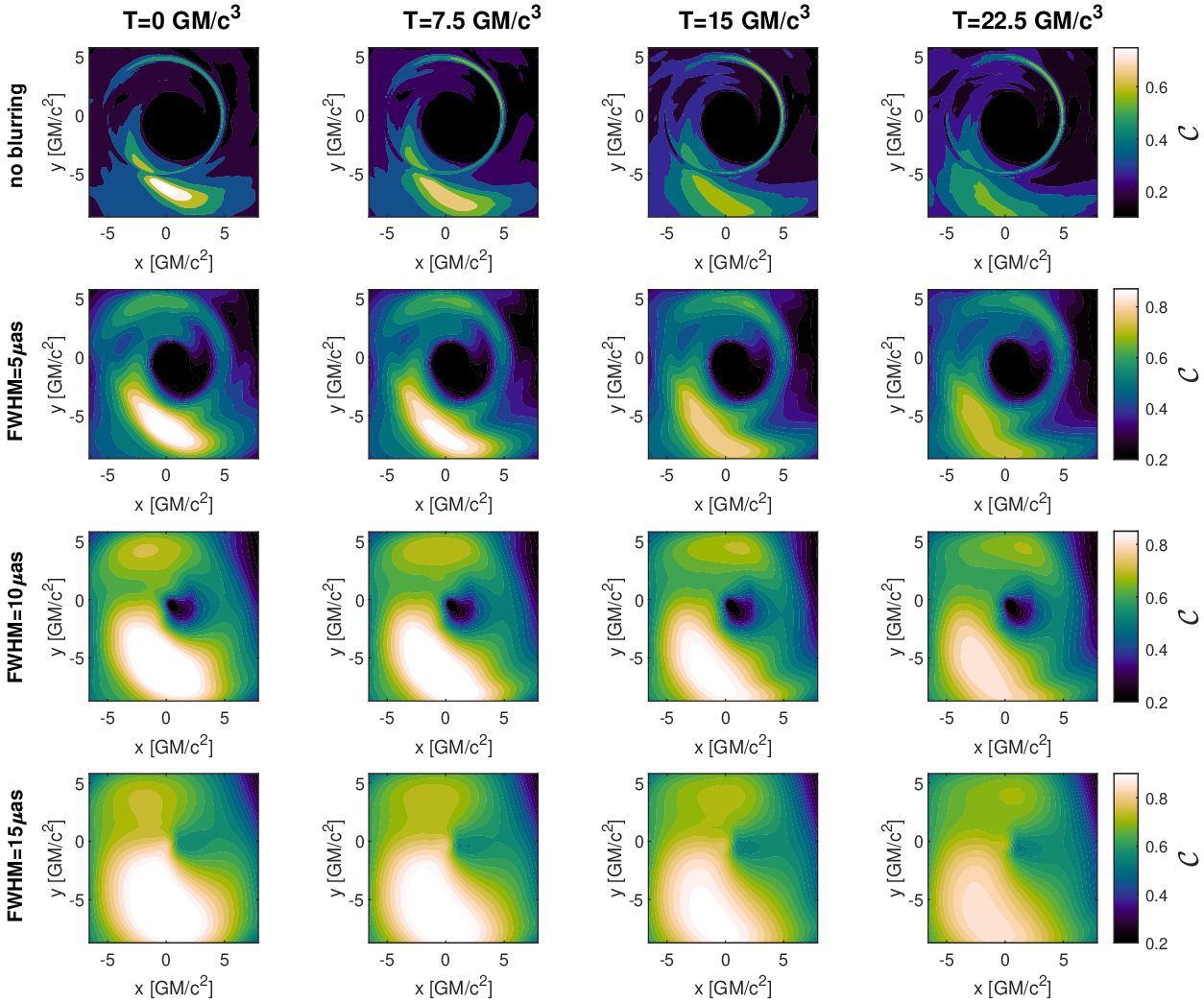} %corr_T_50_20
\caption{Correlation maps $\mathcal{C}(T_0,x,y,x_0',y_0')$ for different choices of time lag $T_0$ and blurring kernel widths for a fixed point at $(x_0',y_0')=(-0.132, -6.190)~GM/c^2$. Each map was generated from slow-light ray-traced movies. The astrophysical correlation lobe peaks in the bottom middle part of the image at $T=0 \, GM/c^3$, while the lensing correlation lobe peaks near the upper middle area of the image around $T=15 \, GM/c^3$.} %These correlation maps are consistent with the presence of a local maximum of $\mathcal{C}(T,x,y,x_0',y_0')$ in the 3-dimensional configuration space $\{T,x,y\}$, constituting a signature of extreme lensing.
\end{figure*}

\begin{figure*}[ht!]
\centering
\includegraphics[width=1\textwidth]{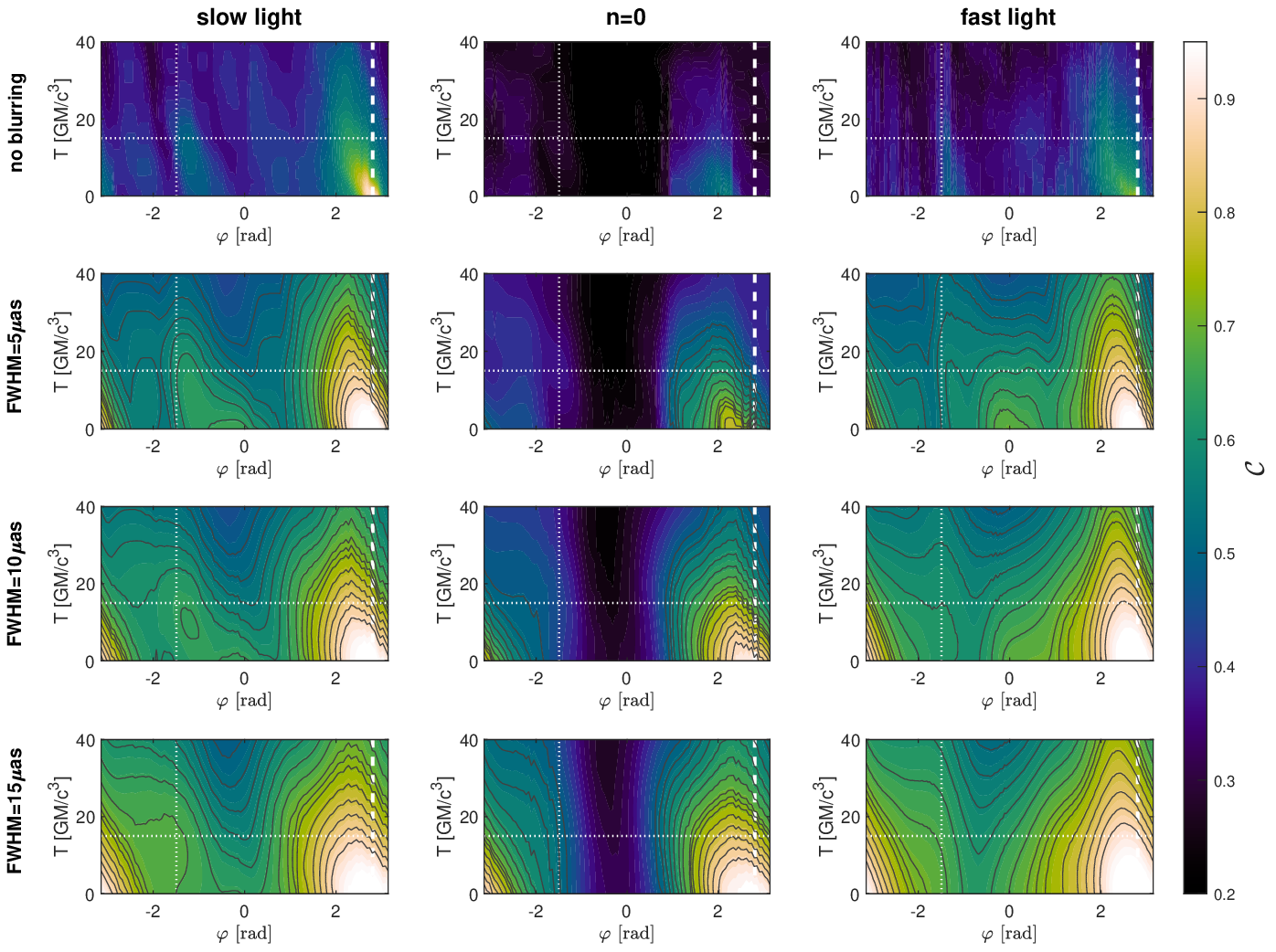} %corr_map_10_80
\caption{Two-point correlation function of intensity fluctuations
as a function of the angular position $\varphi$ and the time lag $T$.
The correlation function $\mathcal{C}(T,\rho_c(\varphi),\varphi,x_0',y_0')$ (see Eq.~1 in the Methods Section) was obtained with different blurring kernels (rows) and ray-tracing prescriptions (columns). The dashed line shows the angular position $\varphi_0'\approx 0.9\pi \, \mathrm{rad}$ of the point which we fixed, located at $(x_0',y_0')=(-5.400, 1.712)~GM/c^2$, and the horizontal and vertical dotted lines mark the expected positions of the secondary peak in the temporal and angular domains, respectively. The contour levels plotted in gray correspond to values 0.5, 0.53, 0.56, 0.59, 0.62, 0.65, 0.68.}
\end{figure*}

\begin{figure*}[ht!]
\centering
\includegraphics[width=1\textwidth]{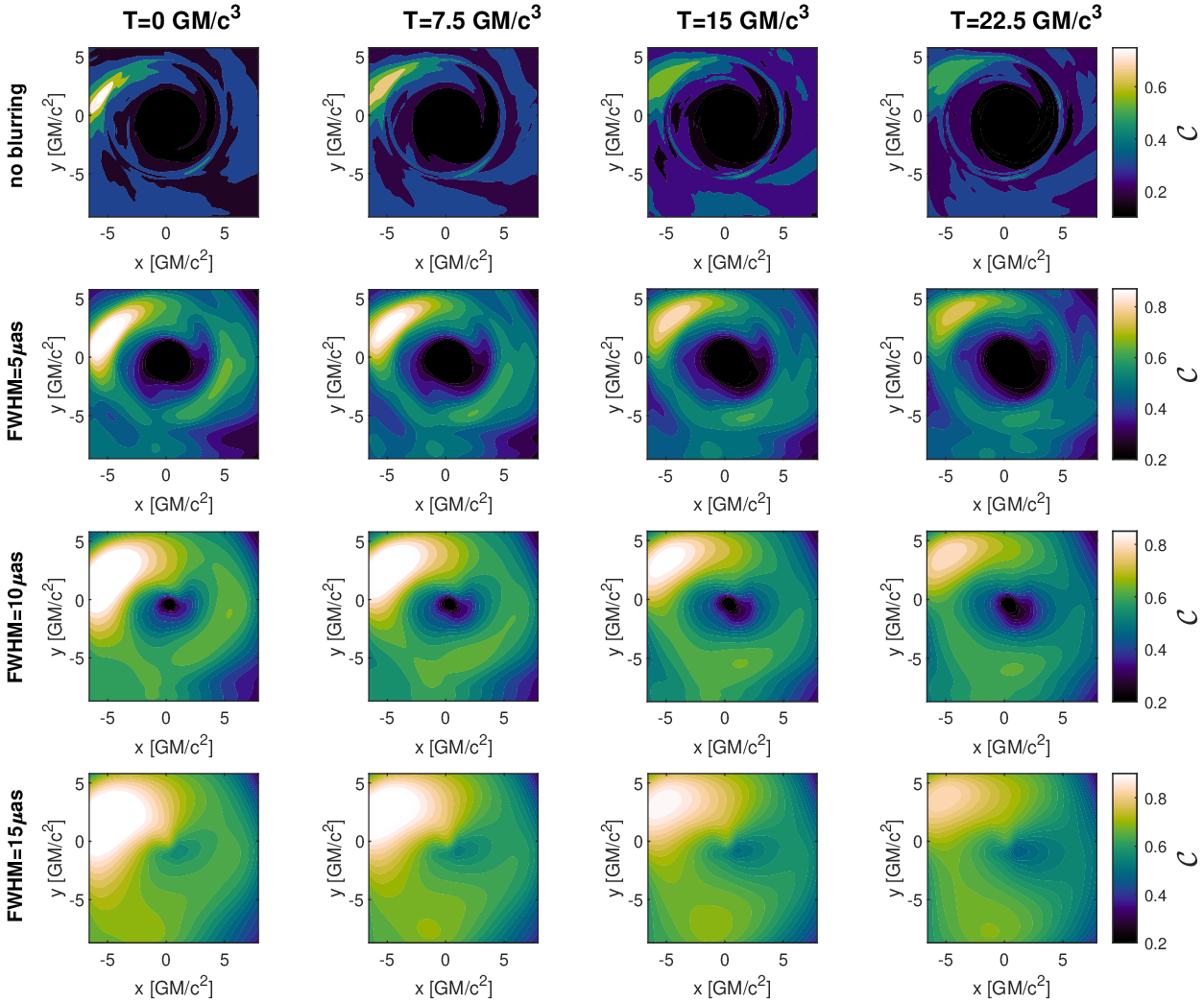}%corr_T_10_80
\caption{Correlation maps $\mathcal{C}(T_0,x,y,x_0',y_0')$ for different choices of time lag $T_0$ and blurring kernel widths for a fixed point at $(x_0',y_0')=(-5.400, 1.712)~GM/c^2$. Each map was generated from slow-light ray-traced movies. The astrophysical correlation lobe peaks near the left middle area of the image at $T=0 \, GM/c^3$, while the lensing correlation lobe peaks in the bottom middle part of the image around $T=15 \, GM/c^3$.}\label{last supplementary map}
%These correlation maps are consistent with the presence of a local maximum of $\mathcal{C}(T,x,y,x_0',y_0')$ in the 3-dimensional configuration space $\{T,x,y\}$, constituting a signature of extreme lensing.
\end{figure*}

\end{document}